\documentclass[aps,prl,preprint,superscriptaddress,showpacs]{revtex4-1}
\usepackage{graphicx}
\usepackage{color}
\usepackage{amssymb}

\begin{document}

\title{Microscopically-resolved simulations
  prove the existence of soft cluster crystals}

\author{Dominic A. Lenz} \affiliation{Faculty of Physics, University
  of Vienna, Boltzmanngasse 5, A-1090 Vienna, Austria}

\author{Ronald Blaak} \affiliation{Faculty of Physics, University of
  Vienna, Boltzmanngasse 5, A-1090 Vienna, Austria}

\author{Christos N.\ Likos} \affiliation{Faculty of Physics,
  University of Vienna, Boltzmanngasse 5, A-1090 Vienna, Austria}

\author{Bianca M. Mladek} \affiliation{Department of Structural and
  Computational Biology, Max F. Perutz Laboratories GmbH, University
  of Vienna, Campus Vienna Biocenter 5, A-1030 Vienna, Austria}

\date{\today}

\begin{abstract}
We perform extensive monomer-resolved computer simulations of
suitably-designed amphiphilic dendritic macromolecules over a broad
range of densities, proving the existence and stability of cluster
crystals formed in these systems, as predicted previously on the basis
of effective pair potentials [B.\ M.\ Mladek {\it et al.},
  Phys.\ Rev.\ Lett.\ {\bf 96}, 045701 (2006)]. Key properties of
these crystals, such as the adjustment of their site occupancy with
density and the possibility to heal defects by dendrimer migration,
are confirmed on the monomer-resolved picture. At the same time,
important differences from the predictions of the pair potential
picture, stemming from steric crowding, arise as well and they place
an upper limit in the density for which such crystals can exist.
\end{abstract}
\maketitle

Soft cluster crystals (also termed bubble solids), i.e.,~solids where
each lattice site is occupied by several, interpenetrating particles,
arise in systems governed by generic, {\it purely repulsive}
interactions which remain {\it finite} even if particles fully overlap
\cite{Klein94,Likos01,Mladek06}. The completely different character of
these novel cluster crystals, as opposed to hard cluster
phases~\cite{Sci04,Sedgwick05}, is testified by the fundamentally
distinct nature of the interactions between their constituent
particles. The clusters are not stabilized by a combination of
short-range attractions and long-range repulsions, which result in
well-defined aggregates in the liquid that disappear at high
densities~\cite{Sci04,Sedgwick05}.  Rather, clustering is here a
self-enhancing, cooperative phenomenon, where full overlaps with few
particles are favored over costly partial overlaps with many
neighbors. Thus, clusters remain stable due to their mutual
repulsions with the surrounding ones~\cite{Mladek08a}.  
Cluster crystals are thus high-density phases, in which fuzzy
agglomerates, already formed in the liquid, freeze into well-defined
and spatially ordered groups at higher concentrations
\cite{Mladek06}. Predicted for certain
macromolecular~\cite{Mladek08,Narros10,Lenz11} or
atomic~\cite{Goerbig03,Cooper08,Saccani11,saccani:prl:2012} systems,
such clusters form exotic crystals that show mass
transport~\cite{Moreno07,Coslovich11}, unusual reaction to
compression~\cite{Mladek06,Mladek07b,Mladek08a} and
shear~\cite{Nikoubashman12}, and rich phase behavior with a cascade
of isostructural transitions~\cite{Neuhaus11, Zhang10}.  Further, they
possess a unique rheological behavior, showing thixotropy and flow
quantization~\cite{Nikoubashman12}. 

All hitherto performed studies of the properties of soft cluster
solids have been based on the assumption of ultrasoft, {\it pairwise
  additive}, {\it density-independent} inter-particle
interactions. According to theory, the necessary requirement for a
system to display clustering behavior is that its pair interaction
shows negative Fourier components~\cite{Likos01,Likos07}. The pair
potential picture is realistic when understood as an effective
representation of certain types of polymeric macromolecules in {\it
  not too concentrated} solutions
\cite{Dautenhahn94,Louis00,Capone10}.  However, the relevance of such
an approach for experiments is here put into question since the
formation of clusters is a high density phenomenon. Under such
conditions, many-body effects, conformational deformations and
crowding are expected to play a crucial role, which could either
hinder or favor cluster formation.  For ring polymers, the
zero-density pair interaction has negative Fourier components;
however, due to the shrinkage of the rings at finite densities, the
clustering ability is lost~\cite{Narros10}. In this Letter, we show on
a microscopic basis that soft cluster crystals can indeed be
stabilized in a broad density regime for suitably designed
macromolecular systems, motivating the experimental realization of
such systems.

Amphiphilic dendrimers constitute such a system predicted to show
clustering \cite{Mladek08,Lenz11}.  Here, we consider amphiphilic
dendrimers of second generation with two central monomers. The
dendrimers' 14 monomers are divided into two classes: the eight
outermost monomers form a solvophilic shell which surrounds the
solphophobic core build from the interior generations' monomers. At sufficiently close separations, all
monomers strongly repel each other via Morse potentials, while bonds
between monomers are modeled by FENE springs~\footnote{See
  Supplemental Material for details on the
  dendrimer model chosen here.}.  In previous studies, the
zero-density effective pair interaction between various realizations
of such dendrimers has been determined~\cite{Mladek08}, showing the
desired negative Fourier components~\footnote{See Supplemental
  Material (Fig. S1) for the general shape of
  the effective interactions of amphiphilic dendrimers.}. Indeed, we
then found the onset of clustering in the {\it fluid} phase of {\it
  microscopically-resolved} amphiphilic dendrimers~\cite{Lenz11}. Here
we demonstrate the stability of the cluster solid phase (see scheme in
Fig.~S1 of the Supplemental Material). Let $\rho = N/V$ be the density
of a system of $N$ dendrimers in the volume $V$ and $R_g$ be the
$\rho=0$ radius of gyration of such a dendrimer.  An estimate,
$\tilde\rho_f$, of the freezing density $\rho_{f}$ can be obtained on
the basis of the infinite dilution pair potential $\Phi_{\rm
  eff}(r)$~\cite{Likos07}. Further, theory predicts the number of
dendrimers per cluster, i.e., the occupancy $\tilde N_{\text{occ}}$,
to be proportional to the density: $\tilde N_{\rm occ} \propto
\rho$~\cite{Likos01,Mladek06}.  Consequently, the lattice constant
$\tilde a \propto (\tilde N_{\rm occ}/\rho)^{1/3}$ is predicted to be
density-independent. For the dendrimers considered here, lattice site
occupancies of $\tilde N_{\text{occ}} = 3.5$ (bcc) and $\tilde
N_{\text{occ}} = 3.2$ (fcc) are predicted at the freezing density
$\tilde \rho_f R_g^3 = 0.281$. The lattice constants are predicted as
$\tilde a = 2.9R_g$ (bcc) and $\tilde a = 3.6R_g$ (fcc), leading in
both cases to nearest-neighbor distances $\tilde d = 2.5R_g$.

\begin{figure}[tb]
  \includegraphics[width=8.5cm]{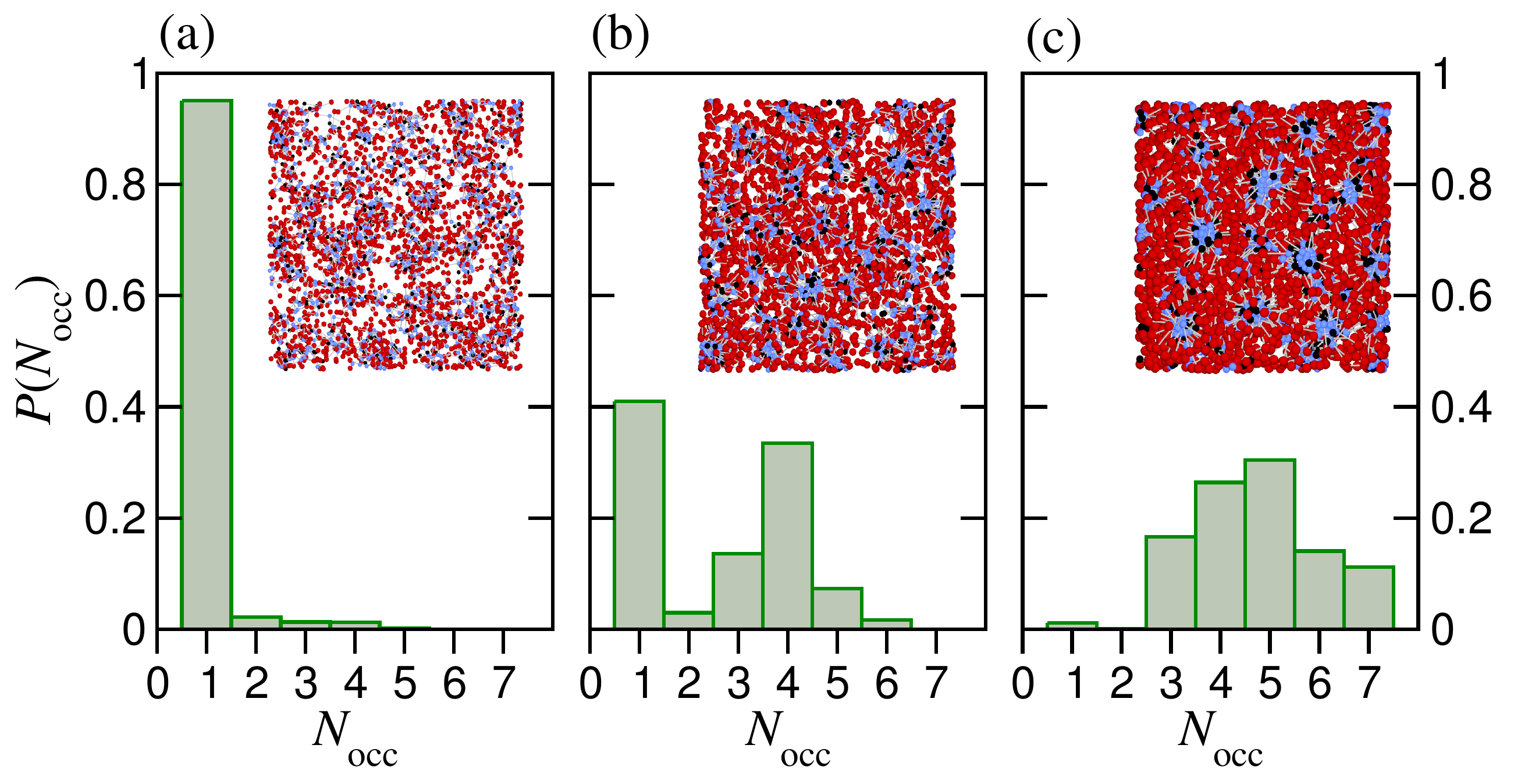}
  \caption{Color online. The probability distribution
    $P(N_{\text{occ}})$ to find cluster size
    $N_{\text{occ}}$ in (a) the fluid state at $\rho R_g^3=0.046$; (b)
    the cluster fluid state at $\rho R_g^3=0.138$; and (c) the solid
    state at $\rho R_g^3=0.323$. The insets show simulation snapshots,
    where clustering can be seen from the crowding of the dendrimers'
    core monomers [black and blue (light gray)]. Shell monomers are shown in red (dark grey). Monomers are not drawn to scale.}
  \label{F:sizeDistribution}
\end{figure}

These theoretical predictions are put to the test by performing
NVT-Monte Carlo simulations. First, we delimit the range of values of
$N_{\rm occ}$ that can stabilize crystals: although in the effective
description $N_{\rm occ}$ can be arbitrarily large, in the
monomer-resolved description local steric crowding is anticipated to
put an upper limit to this quantity. To this end, we simulate {\it
  disordered} systems of dendrimers in a density regime around $\tilde
\rho_f $. We study the cluster size probability distribution
$P(N_{\text{occ}})$, employing a cluster algorithm developed
previously \cite{Lenz11}. Representative results are shown in
Fig.~\ref{F:sizeDistribution} for three different
densities. Fig.~\ref{F:sizeDistribution}(a) shows $P(N_{\text{occ}})$
of a system at low density, $\rho R_g^3 = 0.046$, i.e., well below
$\tilde \rho_{f} $. Some small clusters are present but single
dendrimers dominate: this system is a fluid of single
dendrimers. Increasing the density to $\rho R_g^3=0.138$,
Fig.~\ref{F:sizeDistribution}(b), while still staying below $\tilde
\rho_f $, we find that a distinctive distribution of larger clusters
has also developed. The sizes of these clusters are not uniformly
distributed; rather a single cluster size, $N_{\rm occ}^{\rm pref}$,
is favored, in full agreement with theoretical
predictions~\cite{Likos07}. Here, the system forms a cluster fluid, as
already studied in detail in previous work, albeit for a different
dendrimer model only showing clustering at higher occupancies than the
present model~\cite{Lenz11,LenzXX}.  Finally, the density was
increased above $\tilde \rho_f$ to the value $\rho R_g^3 = 0.323$,
Fig.~\ref{F:sizeDistribution}(c). Now, single dendrimers have almost
completely vanished from the solution and have been replaced by
clusters of a preferred occupation of $N_{\text{occ}}^{\rm
  pref}=4-5$. Visual inspection reveals that this system has
spontaenously developed local crystalline order, albeit forming a
distorted crystal. This finding is corroborated by an analysis of the
radial distribution function of the clusters' centers of mass, which
has developed a pronounced first peak centered at a value close to the
predicted $\tilde a$ and clearly separated from the second
peak. System sizes, however, do not allow for a positive determination
of a particular structure. Similar results were found for several
other densities above $\tilde \rho_f $, proving that the cluster
liquid is {\it unstable} in this density regime.

\begin{figure}[tb]
  \includegraphics[width=8.5cm]{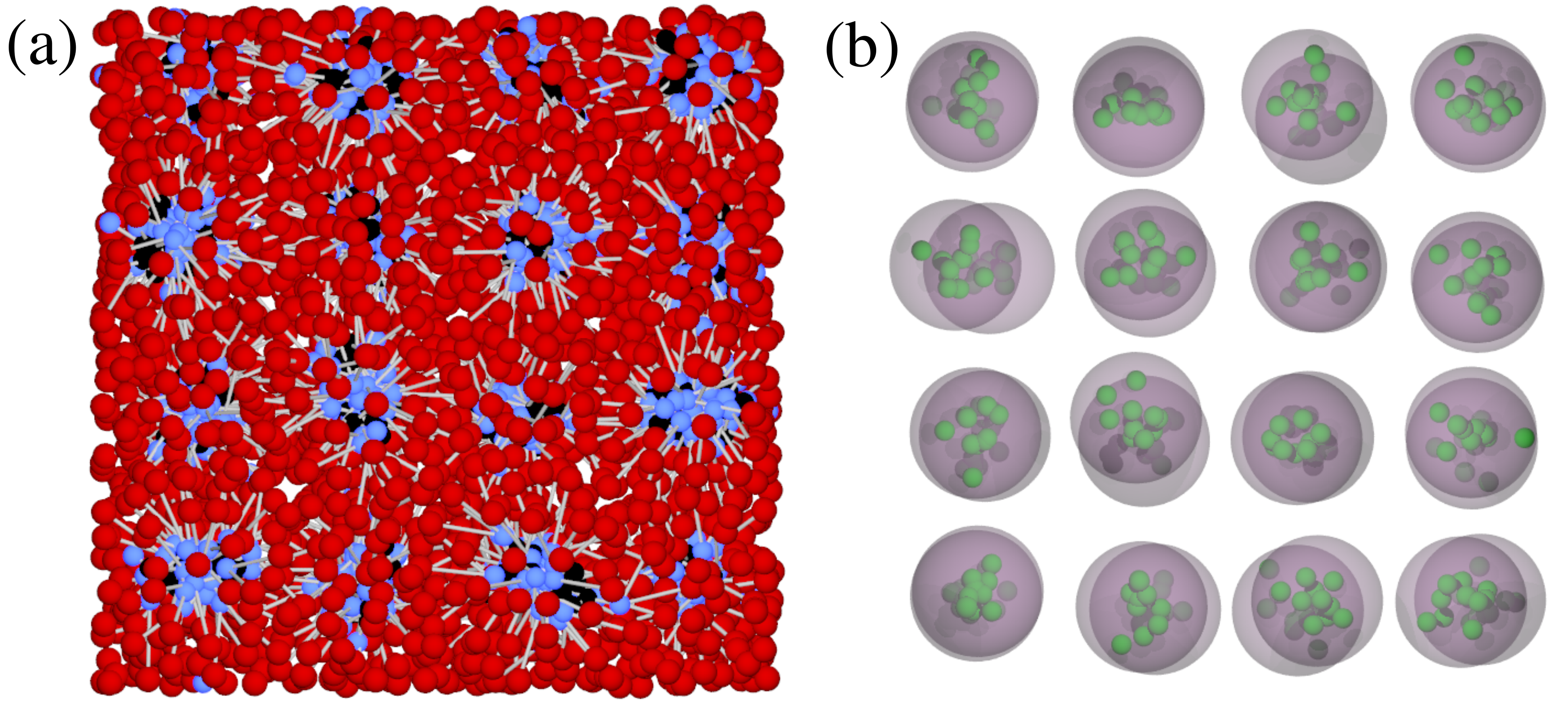}
  \caption{Color online. Simulation snapshot of a fcc dendrimer cluster crystal of
    occupancy $N_{\text{occ}}=10$ at $\rho R_g^3=0.438$.  (a) A
    full-monomer representation of the crystal. Clustering can be seen
    from the crowding of core monomers [black and blue (light gray)].
    (b) The dendrimers' centers of mass (small spheres) are found to
    99$\%$ in the big, gray spheres around the cluster's centers of mass.}
  \label{F:snapshots}
\end{figure}

Spontaneous crystallization, however, does not provide information
which crystalline structure is the one of lowest free energy;
crystallites formed could exhibit the structure with the lowest
barrier separating it from the fluid state
\cite{Alexander78,Ostwald84}.  We therefore artificially prepared
perfect bcc and fcc cluster crystals at various densities $\rho$, by
placing $N_{\text{occ}} \gtrsim N_{\rm occ}^{\rm pref}(\rho)$
dendrimers per perfect lattice site and then delimited their range of
mechanical stability by performing NVT- and NPT-Monte Carlo
simulations.  Non-stable crystals melted within a few simulation
sweeps into diffusive fluids, as manifested by a steady increase of
the dendrimers' mean-square displacement and an accompanying loss of
the ordered structure. The mechanically stable systems, however,
remained in the crystalline, clustered arrangement throughout the
whole simulation runs. Fig.~\ref{F:snapshots}(a) shows a typical
simulation snaphot of a cluster crystal in the full-monomer
representation of the dendrimers, where the clustering can be seen
from the crowding of core monomers [black, blue (light gray)], which
are surrounded by a cloud of shell monomers [red (dark gray); also
  cf.~Fig.~S1].  The clustering and the self-assembly of the clusters
onto the lattice sites of a (fcc) crystal become even more evident
when plotting just the dendrimers' centers of mass,
Fig.~\ref{F:snapshots}(b), denoted there by small spheres.  The
irregular shapes of the various clusters are a manifestation of the
instantaneous fluctuations of each dendrimer around its lattice site
and show that the centers of mass of the individual dendrimers are not
tightly bound to lie on top of each other. These findings are in full
agreement with the theoretical picture, which predicts these
oscillations to give rise to intra-cluster breathing modes that
manifest themselves as optical branches in the phonon spectrum of the
crystal \cite{Neuhaus11}.  We quantify these fluctuations by plotting
gray spheres, centered around the clusters' centers of mass, and which
contain the latter to a probability of 99$\%$. Spheres from different
lattice sites do not overlap, confirming the prediction that the
clusters are well-localized around their lattice sites \cite{Likos07}.

\begin{figure}[tb]
  \includegraphics[width=8.5cm]{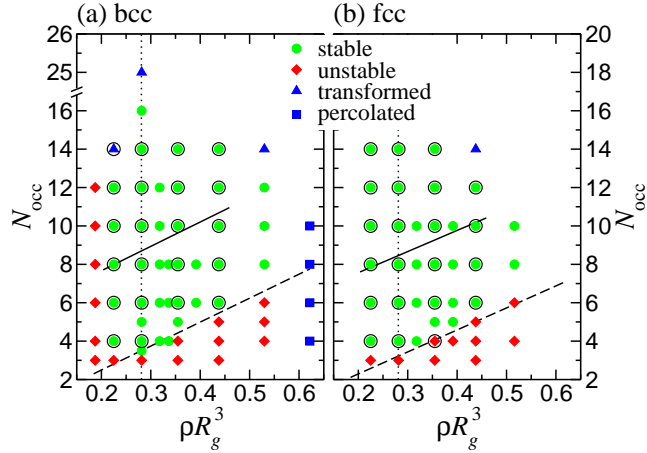}
  \caption{Color online. The mechanical stability of dendrimer cluster
    crystals for (a) bcc and (b) fcc as function of density $\rho
    R_g^3$ and occupation number $N_{\rm occ}$. State points for which
    crystals remained stable are denoted by circles ($\bullet$), those
    where crystals melted as diamonds ($\blacklozenge$). The predicted
    freezing density $\tilde\rho_f$ is shown as vertical, dotted line at $\rho R_g^3
    = 0.281$. For high occupations and densities, systems transformed
    ($\blacktriangle$) or formed percolated networks ($\blacksquare$)
    (see text). Systems for which the free energy was measured are
    circled in black. The dashed line plots the theoretical prediction
    $\tilde N_{\rm occ}(\rho)$, while the equilibrium occupation
    $N_{\rm occ}^{\rm eq}(\rho)$ (see text) is plotted as solid line.}
  \label{F:mechanicalstability}
\end{figure}

The results of the simulations delimiting the range of mechanical
stability are summarized in Fig.~\ref{F:mechanicalstability},
revealing that a large, insular part of density-occupancy space allows
for the formation of mechanically stable cluster solids. The minimum
density to provide mechanical stability of the cluster crystals is
found to be $\rho R_g^3 \cong 0.225$, independently of the occupation
number. While the determination of the precise location of the
freezing density $\rho_f$ is not the subject of this paper, these
results demonstrate that $\rho_f \simeq \tilde \rho_f$.

As can be seen in Fig.~\ref{F:mechanicalstability}, the minimum
occupation number leading to mechanically stable solids increases
linearly with density for both bcc and fcc. Systems with the lowest
stable occupation, $N_{\text{occ}}\simeq 4$ showed single events of
lattice site hopping and subsequent healing of the resulting
defect. However, due to the limited time-scales accessible in our
simulations and the crowded systems studied, complex structural
rearrangements and hopping events leading to long-time diffusion
\cite{Moreno07,Coslovich11} have not been observed frequently.
Crystals of high occupations, $N_{\rm occ} \simeq 14$, further
remained mechanically stable upon artificially introducing lattice
defects like single clusters of lower occupancy, or even lattice
vacancies or interstitials. Also mixed occupations, where not
every lattice site was occupied by the same amount of dendrimers,
remained stable, provided the average value of $N_{\rm occ}$ was
chosen from the island of stability (see
Fig.~\ref{F:mechanicalstability}), in full agreement with the
theoretical prediction of cluster polydispersity in the crystal
\cite{Likos07,Moreno07}.

Crystals with even higher occupations, i.e., $N_{\text{occ}}\gtrsim
16$ were not found to be stable: they transformed to distorted crystal
structures of lower occupations by splitting clusters and halving the
lattice constant, thereby taking these crystals back to the island of
stability (Fig.~\ref{F:mechanicalstability}, triangles). This shows
the propensity of over-occupied crystals to adjust their occupancy
towards an optimal value, a feature which will be confirmed
quantitatively by free energy calculations in what follows. At high
densities and rather low occupations, the lattice spacing accordingly
shrinks and enters the length regime of the bonds between the
dendrimers' monomers. Here, the crystalline arrangement of
well-defined clusters also cannot be sustained: dendrimers stretch and
are shared between two clusters, leading to disordered percolated
networks of dendrimers (Fig.~\ref{F:mechanicalstability},
squares). This phenomenon reflects that at such high densities the
effective pair potential picture, which predicts an unlimited growth
of cluster occupancy with density, $N_{\rm occ} \propto \rho$, breaks
down {\it qualitatively} since steric crowding between the monomers
prevents further cluster growth upon sufficient compression.

\begin{figure*}[tb]
  \includegraphics[width=8.5cm]{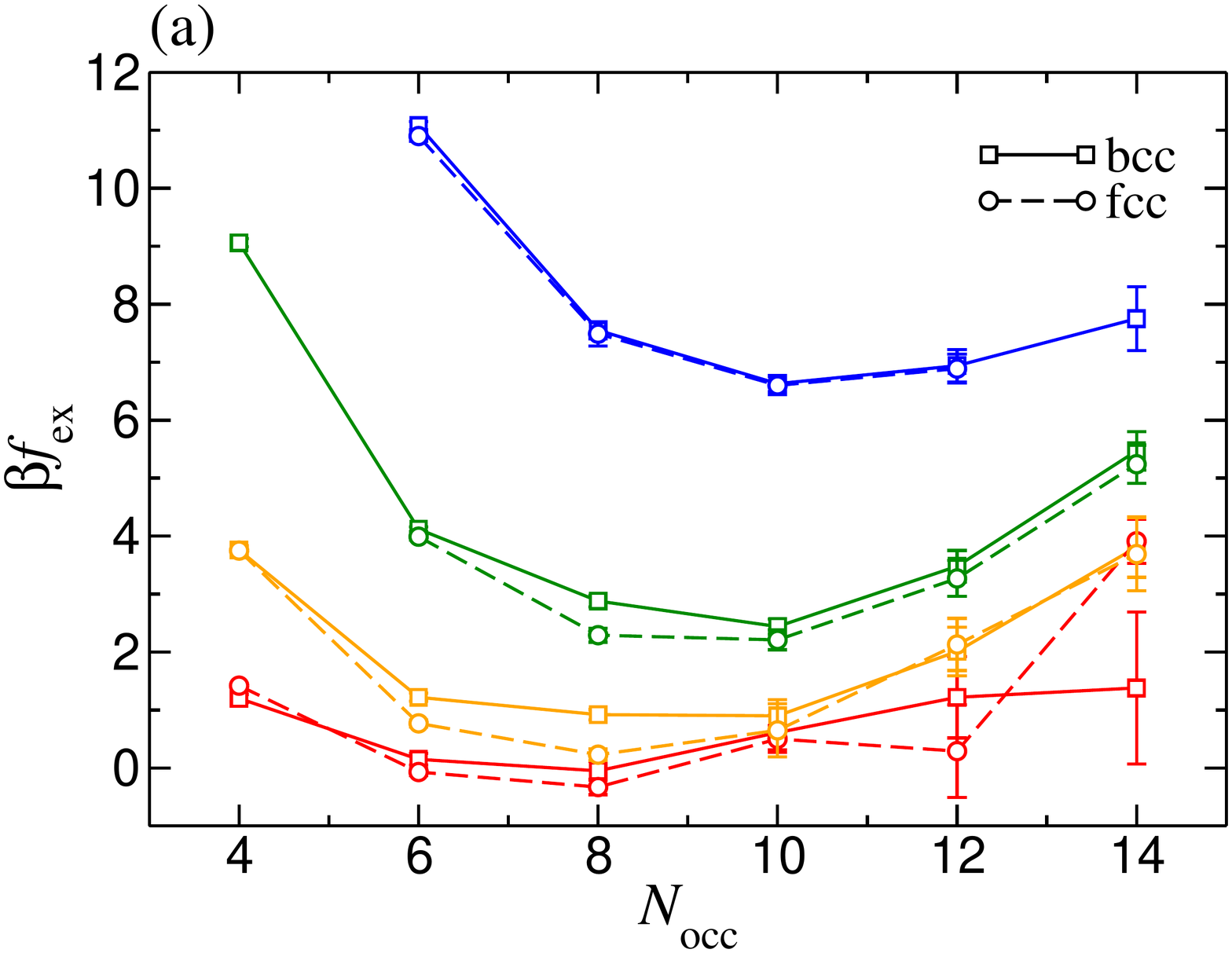}
  \includegraphics[width=8.5cm]{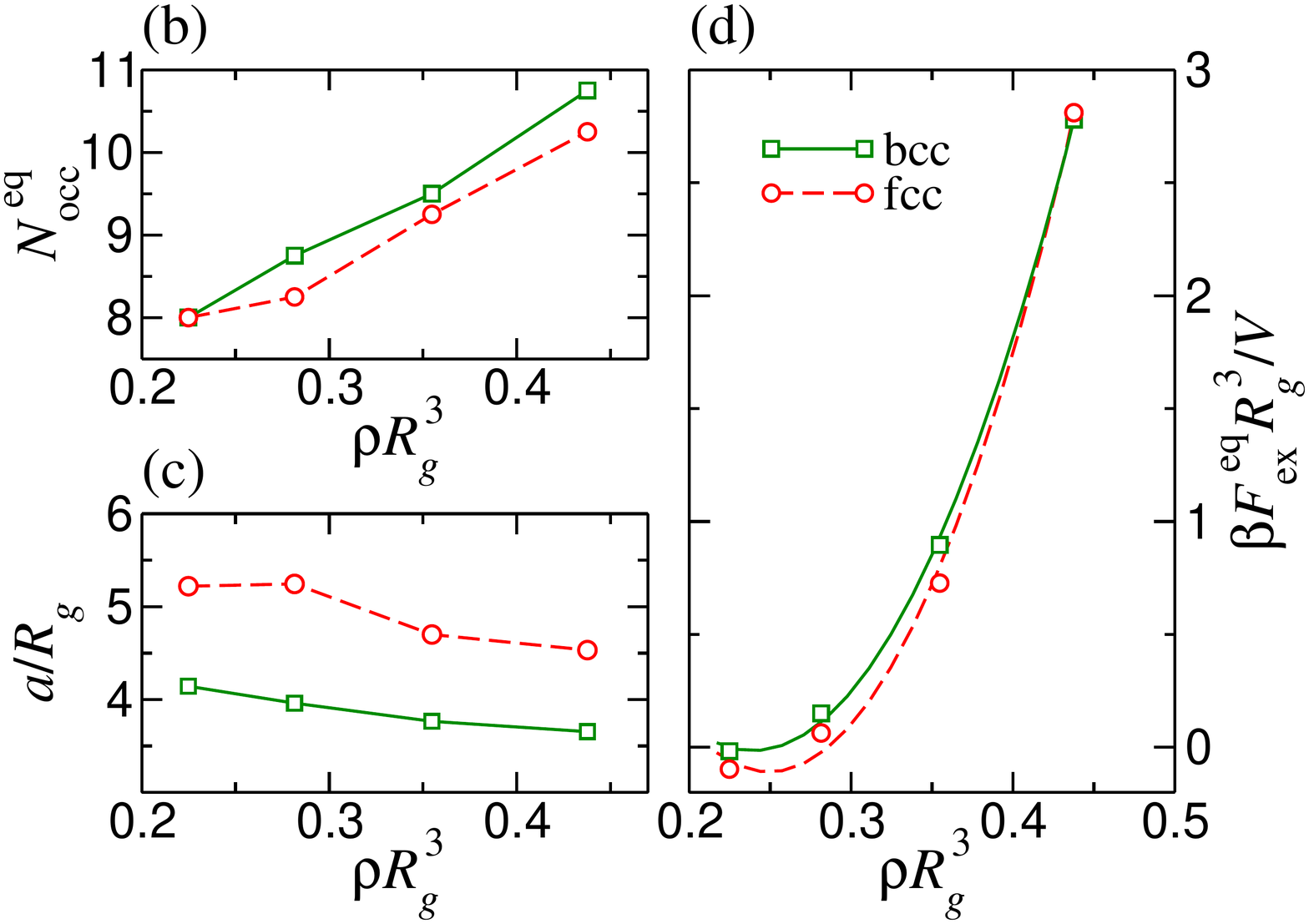}
  \caption{Color online. All data is plotted for fcc ($\circ$) and bcc
    ($\Box$) lattices. (a) The free energy per dendrimer, $\beta
    f_{\rm ex}$, as a function of $N_{\rm occ}$ at different
    densities, from bottom to top: $\rho R_g^3 = 0.22$, $0.28$, $0.36$
    and $0.44$, showing for each density a different optimal
    occupancy, $N_{\rm occ}^{\rm eq}$, that minimizes it. (b) The
    equilibrium occupancy, $N_{\rm occ}^{\rm eq}$, as function of
    density $\rho R_g^3$. (c) The equilibrium lattice constant $a/R_g$
    shrinks with increasing density. (d) The equilibrium free energy
    per volume, $\beta F_{\text{ex}} R_g^3/V$ as a function of the
    density $\rho$, evaluated at the equilibrium occupancy $N_{\rm
      occ}^{\rm eq}(\rho)$. Fcc is found to be the thermodynamically
    stable crystal.}
\label{F:FreeEnergyResults}
\end{figure*}

Having established the existence of a broad region of their {\it
  mechanical} stability, it is pertinent to assess the crystals' {\it
  thermodynamic} stability and their optimal occupation at given
density. To this end, we calculated the excess free energy per
particle $\beta f_{\text{ex}} = \beta F_{\text{ex}}/N$ for the
mechanically stable crystals with the help of thermodynamic
integration as outlined in Refs.~\cite{Mladek07b,Mladek08a}. Here, the
reference state is chosen as non-interacting dendrimers confined to
the desired crystal structure by potential wells. Being the same for
all crystal structures, the reference free energy $\beta F_{\rm id}/N$
may be disregarded. Measuring $\beta f_{\text{ex}}$ for several values
of $N_{\text{occ}}$ at given crystal structure and fixed density, we
determine the equilibrium occupancy $N_{\rm occ}^{\rm eq}$ as the one
that minimizes the free energy, see Fig.~\ref{F:FreeEnergyResults}(a)
\cite{Zhang10}. In qualitative agreement with theoretical predictions,
$N_{\rm occ}^{\rm eq}$ increases linearly with density,
Fig.~\ref{F:FreeEnergyResults}(b).  However, there exist quantitative
deviations between the two, manifesting the role of the many-body
interactions. Whereas the pair-potential description predicts $N_{\rm
  occ} \propto \rho$, here there is a non-zero offset $d$, $N_{\rm
  occ} = c \rho + d$. At freezing, the optimal occupation number from
the monomer-resolved simulations turns out to be {\it higher} than
theoretically predicted. This indicates that the many-body
interactions, whose importance grows with occupancy, have the effect
of strengthening the overall attractions between the soft particles at
close approach. Since these properties are model-specific, we do not
attempt a more detailed analysis here. However, we point out that the
theoretical prediction of $\tilde N_{\rm occ}(\rho)$ is found to be an
excellent estimate for the lower limit in occupancy of mechanically
stable crystals (see Fig.~\ref{F:mechanicalstability}). Indeed, these
occupancies are sufficiently low for pair potentials to still be
approximately valid, their degree of validity diminishing as $N_{\rm
  occ}$ grows. In Fig.~\ref{F:FreeEnergyResults}(d) we show the free energy per
volume of the equilibrium crystals as a function of density. We find
that the fcc crystals have a slightly lower free energy than bcc for
all densities considered.  %This is contrasted by the fact that
%spontaneous freezing results in distorted bcc crystals with
%significantly lower occupancy than $N_{\rm occ}^{\rm eq}$. We thus
%conjecture that systems pass intermediate steps in the freezing
%process by forming easily accessible bcc crystals
%\cite{Alexander78,Ostwald84} of low occupation, until eventually
%finding a route to the optimal fcc structure. Such transformations
%were not observed in our simulations, since they exceed the accessible
%time scales.

In contrast to amphiphilic dendrimers, ring polymers cannot form
cluster crystals due to their drastic shrinkage upon compression
\cite{Narros10}. It is therefore pertinent to study the
conformations of dendrimers in the clusters of the thermodynamically
stable crystals. The radius of gyration of the dendrimers shows little
variation with density, due to the necessary balance between the
exterior and interior osmotic pressure of clusters and the 
constraints imposed by the regularly branched dendritic architecture.
Nevertheless, the configurations of the dendrimers change considerably
upon increasing density, a fact manifested by a density-dependence of
the most probable distance of shell monomers from their dendrimer's
center of mass. While shell monomers can backfold to the dendrimer's
core at low densities, the cluster formation at finite densities
increasingly forces the shell monomers to move to the rim of the
molecules \footnote{ See Supplemental Material for details on this behavior.}.

Altogether, the cluster crystals discovered in our study are
characterized by a hierarchy of self-organization at various
interconnected levels. First, the individual dendrimers adjust their
conformations to the thermodynamic conditions; second, the individual
clusters adjust their population and are thus robust against cluster
polydispersity and crystal defects; and finally at the macroscopic
level, the flexibility of the cluster crystals is manifested through
the mechanism of restructuring of non-optimally occupied cluster
crystals, the hopping of dendrimers among crystal sites, and the
adjustment of the associated lattice constant. Discrepancies between
the fully microscopic and the effective potential pictures exist due
to the fact that the strong steric repulsions of the monomers prohibit
the {\it ad libidum} growth of the clusters' population. Still, the
salient features of soft cluster crystals have been confirmed. The
range of stability of these structures is sufficiently broad to render
them into materials of practical as well as fundamental importance.

In summary, we have shown at a microscopic level that cluster crystals
exist in suitably synthesized soft matter systems. More detailed
computational investigations in both the effective and the
microscopical description will be needed to uncloak the whole beauty
of these intriguing systems and understand, for example, the
nucleation of soft cluster crystals or the transformations of
dendrimer crystals at high occupations and densities; this will be the
topic of future investigations.  Our results convey the clear message
that an experimental realization of cluster crystals employing simple
and easy to synthesize dendritic macromolecules is possible. We
thereby provide guidelines to achieve this goal, which will lead to the
production of a new class of materials with most intriguing physical
properties.

We thank Daan Frenkel (Cambridge) for helpful
discussions.  D.A.L. thanks the Wolfgang Pauli Institute.
B.M.M. acknowledges the MFPL VIPS program (funded
by BMWF and the City of Vienna), and EU funding (FP7-PEOPLE CIG-2011
No.~303860). Partial support by the ITN-COMPLOIDS (Grant
Agreement No.~234810) and computer time at the Vienna Scientific
Cluster are gratefully acknowledged.

%\bibliographystyle{apsrev}
%\bibliography{manuscript}

\newpage

\begin{center}
\begin{large}
{\bf Supplemental Material for paper:\\ Microscopically-resolved
  simulations prove the existence of soft cluster crystals}
\end{large}
\end{center}

\begin{center}
Dominic A. Lenz, Ronald Blaak, Christos N. Likos, and Bianca M.Ê Mladek
\end{center}

\setcounter{figure}{0}
\makeatletter 
\renewcommand{\thefigure}{S\@arabic\c@figure} 
\makeatother

\section*{The dendrimer model}
\begin{figure*}[tb]
  \includegraphics[width=17cm]{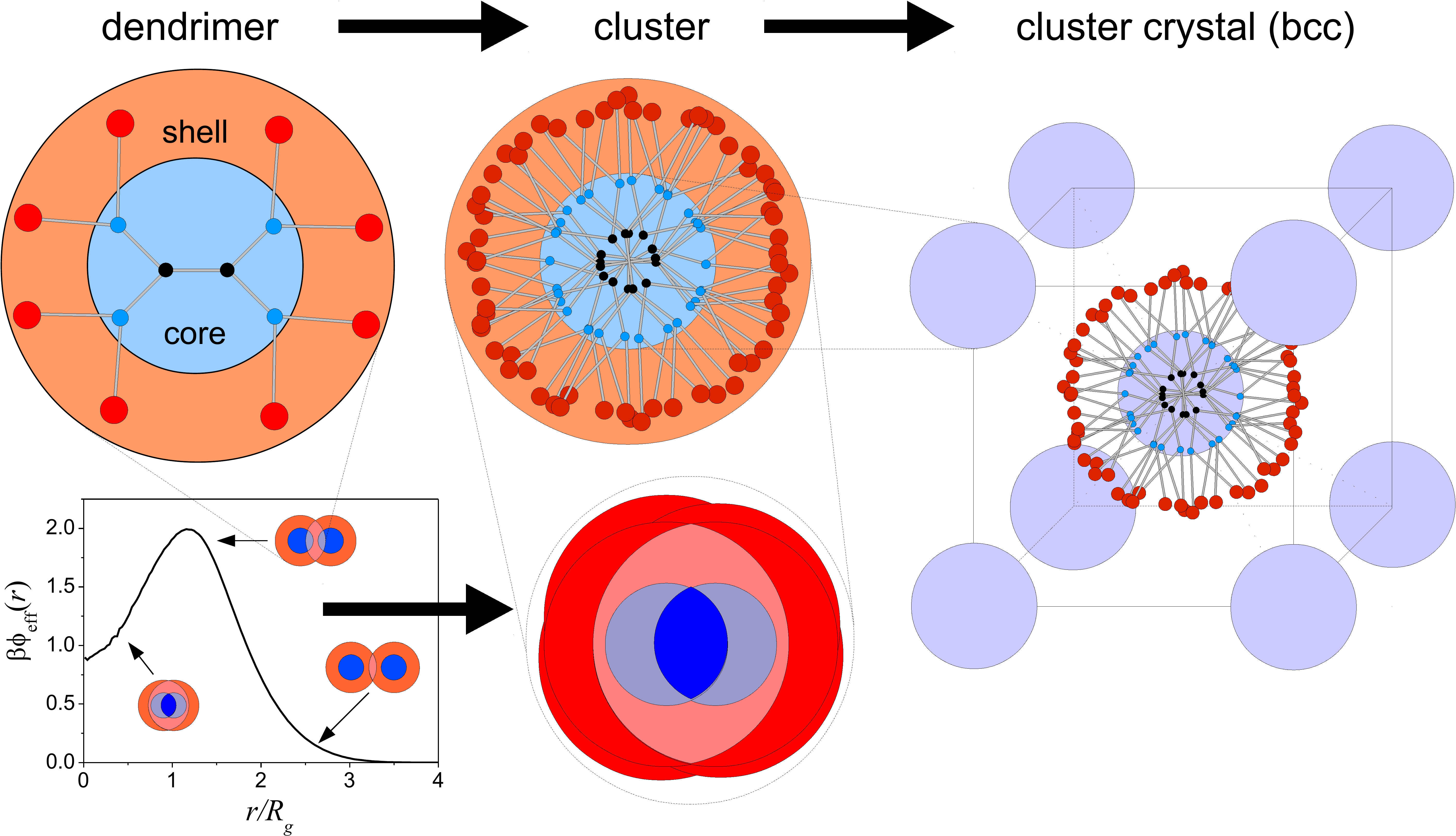}
  \caption{Schematic representation of amphiphilc dendrimers of second
    generation, consisting of a solvophobic core (black and blue inner
    monomers) and a solvophilic shell (red outer monomers; monomers
    not drawn to scale). Several, interpenetrating dendrimers form
    clusters, which is expected to be promoted by the
    dendrimer-dendrimer interaction, $\Phi_{\rm eff}(r)$ (bottom
    left), which favors clustering. Clusters can then freeze into
    crystals. Here, $\beta = (k_BT)^{-1}$ and $R_g$ is the infinite
    dilution radius of gyration.}
  \label{F:scheme}
\end{figure*}

A generic view of the amphiphilic dendrimer architecture, of the
effective pair potential and of the hierarchical assembly of the
molecules into a cluster crystal is shown in Fig.\ \ref{F:scheme}.  We
employ the dendrimer model already introduced before in
Refs.~\cite{Mladek08,Lenz11} and based on Ref.~\cite{Welch98}. We
study amphiphilic dendrimers of second generation which have two
central monomers. The dendrimers' 14 monomers are divided into two
classes: the eight monomers of the outermost generation ($g=2$) form
the solvophilic shell of the dendrimer (index S), and all monomers of
the interior generations ($g=0$ and $g=1$) form the solvophobic core
(index C), resulting in amphiphilic dendrimers. Interactions between
any two monomers, separated by a distance $r$ are modeled by a Morse
potential~\cite{Welch98}, which is given by
\begin{equation}
\beta \Phi_{\mu\nu}^{\text{Morse}} (r) = \epsilon_{\mu\nu} \left\{
  \left[ e^{-\alpha_{\mu\nu}(r-\sigma_{\mu\nu})} - 1 \right]^2 - 1
\right\}, ~~ \mu\nu=\text{CC, CS, SS} 
\label{e:Morse}
\end{equation} 
where $\sigma_{\mu\nu}$ denotes the effective diameter between two
monomers of species $\mu$ and $\nu$. All energy scales employed are
  measured in units of $\beta = (k_{\rm B} T)^{-1}$, with $T$ being
the temperature and $k_{\rm B}$ Boltzmann's constant. The Morse
potential is characterized by a repulsive short-range behavior and an
attractive part at long distances, whose depth and range are
parametrized via $\epsilon_{\mu\nu}$ and $\alpha_{\mu\nu}$,
respectively. The core monomer diameter $\sigma_{\text{c}}$ is chosen
to be the unit of length for the model parameters given in
Tab.~\ref{t:par}.

The bonds between adjacent monomers of relative distance $r$ are
modeled by the finitely extensible nonlinear elastic (FENE)
potential, given by
\begin{equation}
\beta \Phi_{\mu\nu}^{\text{FENE}} (r) = -K_{\mu\nu}R_{\mu\nu}^2 \log
\left[ 1 - \left( \frac{r - l_{\mu\nu}}{R_{\mu\nu}} \right)^2 \right],
~~ \mu\nu =\text{CC, CS}
\label{e:FENE}
\end{equation}
where the spring constant $K_{\mu\nu}$ restricts the monomer
separation to be within a distance $R_{\mu\nu}$ from the equilibrium
bond length $l_{\mu\nu}$. 

We simulated more than 100 different realizations of dendrimers,
determining the zero-density pair interactions via the algorithm
outlined in \cite{Mladek11}. We then chose the dendrimer with the deepest
minimum in Fourier transform, which promises a high ability of
clustering. Its parameters are given in Tab.~\ref{t:par}.

We stress that we also carried out simulations of cluster crystals for
other realizations of dendrimers, whose effective interactions are far
less promising regarding their clustering behavior. While these
systems are more difficult to access computationally due to their
higher necessary equilibrium occupancies, the obtained results are in
full accordance with the findings presented in the main part of the
present contribution.

\begin{table}[th]
\begin{center}
\begin{tabular}{c|l|l|l}
\hline 
\hline
\textbf{Morse} & $\epsilon$ & $\alpha \sigma_{\text{c}}$ & $\sigma/\sigma_{\text{c}}$\\
\hline
\hline
CC & 0.714 & 1.8  & 1.0    \\
\hline
CS & 0.01785 & 6.0  & 1.75 \\
\hline
SS & 0.01785 & 6.0  & 2.5  \\
\hline 
\hline
\textbf{FENE} &  $K\sigma_{\text{c}}^2$ & $l_0/\sigma_{\text{c}}$ & $R/\sigma_{\text{c}}$\\
\hline 
\hline
CC ($g=0$) & 60    & 3.1875  & 0.6375 \\
\hline
CC ($g \neq 0$) & 60    & 1.875 & 0.375\\
\hline
CS & 30    & 3.5625  & 0.7125 \\
\end{tabular}
 \caption{Potential parameters of the dendrimer considered in this
   study. The radius of gyration of an isolated dendrimer is $R_g =
   3.589 \sigma_c$.}
 \label{t:par}
\end{center}
\end{table}

\section*{Dendrimer structure in the cluster crystals}

\begin{figure*}[tb]
  \centering
  \includegraphics[width=8cm]{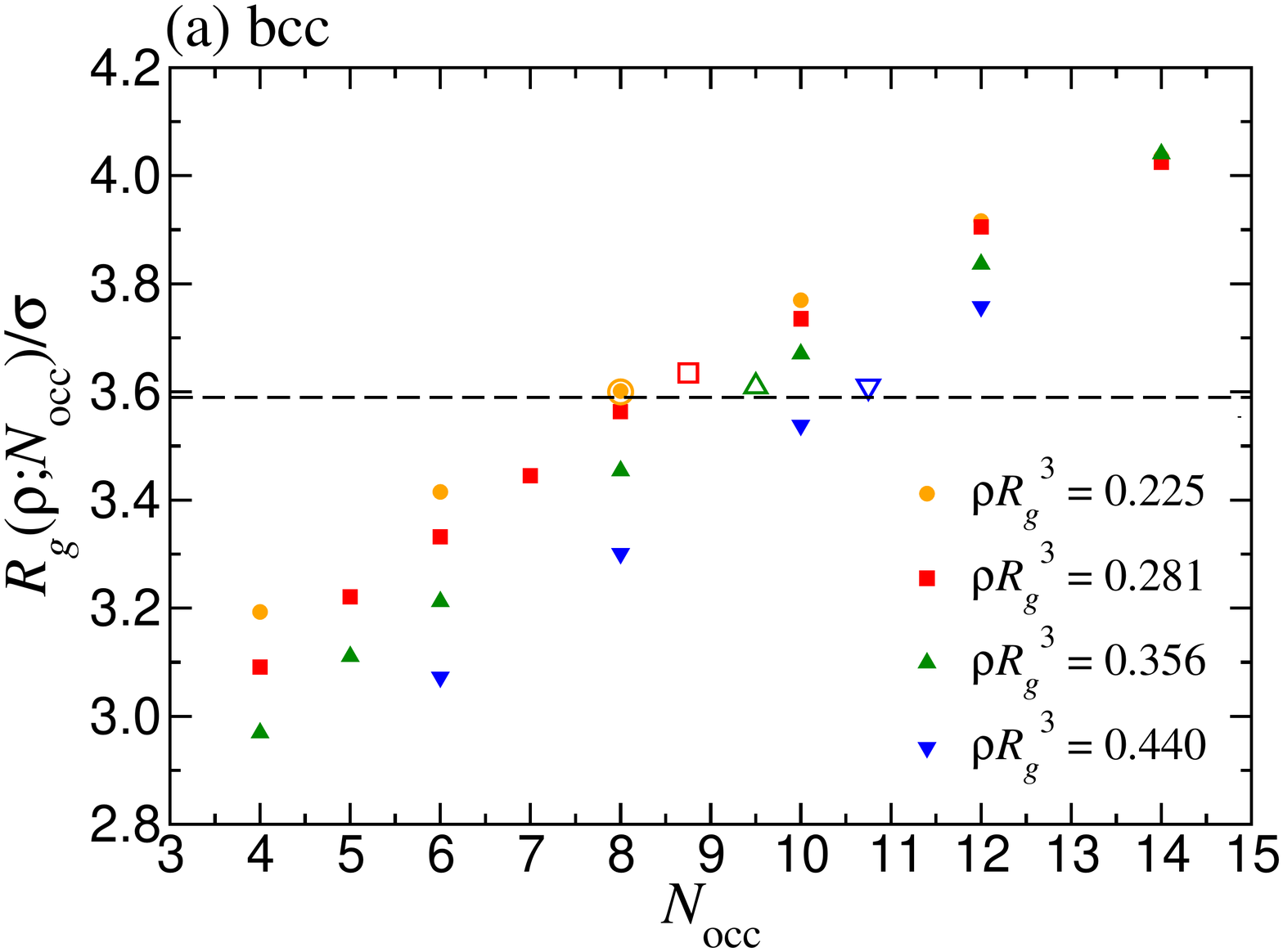}
  \includegraphics[width=8cm]{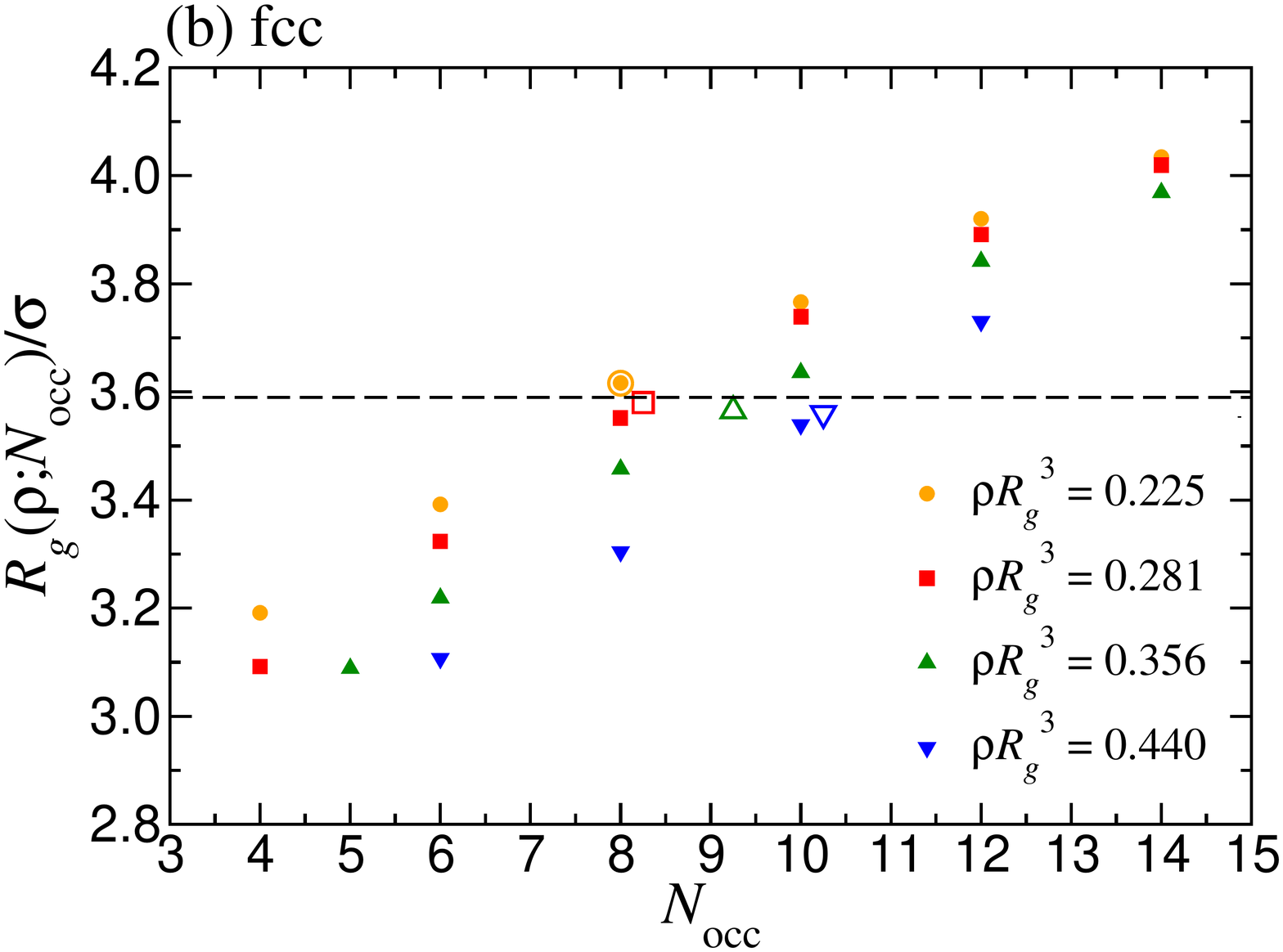}
  \caption{The radius of gyration of (a) bcc and (b) fcc crystals as a
    function of crystal occupation number for various crystal
    densities as indicated in the legends. The open symbols denote the
    value of $R_g(\rho;N_{\rm occ})$ at the optimal occupation number,
    obtained by thermodynamic integration as described in the main
    text, and they are color- and symbol coded as the filled
    symbols. The horizontal line indicates the value $R_g =
    3.59\sigma$ of the gyration radius of the dendrimers at infinite
    dilution.}
  \label{Fig:clustercrystal:RgNocc}
\end{figure*}

In the main text, we refer to the adjustment of dendrimer sizes and
conformations upon changes of the density $\rho$ and the cluster
occupation $N_{\rm occ}$.  Here we show quantitative evidence of this
effect.  The dependence of the dendrimers' radii of gyration on $\rho$
and $N_{\rm occ}$, $R_g(\rho; N_{\rm occ})$, for the mechanically
stable crystals in (cf.~Fig.~3 of main text) is shown in
Fig.~\ref{Fig:clustercrystal:RgNocc}. For reference, the zero-density
value $R_g$ is plotted as well. The equilibrium values of the radii of
gyration $R_g^{\text{eq}}(\rho)$, corresponding to the optimal
occupancy $(\rho,N_{\text{occ}}^{\text{eq}})$ according to
Fig.~4 are displayed by the open symbols,
revealing that their values are nearly density-independent and,
interestingly enough, very close to the zero density value. The
dependence of the dendrimer size on $N_{\rm occ}$ for a fixed density
$\rho$ is a monotonically increasing function, a feature that can be
understood through the influence of two interconnected mechanism. On
the one hand, as $N_{\rm occ}$ grows, the dendrimers have to stretch
to accommodate the increasing number of monomers. On the other, the
concomitant increase of the lattice constant brings the neighbors
further apart and therefore it slows down the growth of the repulsions
from the neighboring dendrimers, which have the tendency of shrinking
the molecules. The insensitivity of the equilibrium value to the
density can be understood in terms of an equality of the osmotic
pressure from the interior of the clusters, $\Pi_{\rm int}$, and that
from the neighboring clusters, $\Pi_{\rm ext}$. As both scale
proportionally to $N_{\rm occ}$ and the lattice constant is rather
insensitive to the density, this quantity cancels out from both sides
of the equation, bringing about the aforementioned insensitivity of
$R_g$ on $\rho$. This rigidity of the dendrimers with respect to
strong deformations lies at stark contrast to the behavior of linear
or ring-shaped polymer chains, which shrink above their overlap
density $\rho^*$ according to the law $R_g(\rho)/R_g =
(\rho/\rho^*)^{-1/8}$.

\begin{figure*}[htbp]
  \centering \includegraphics[width=8cm]{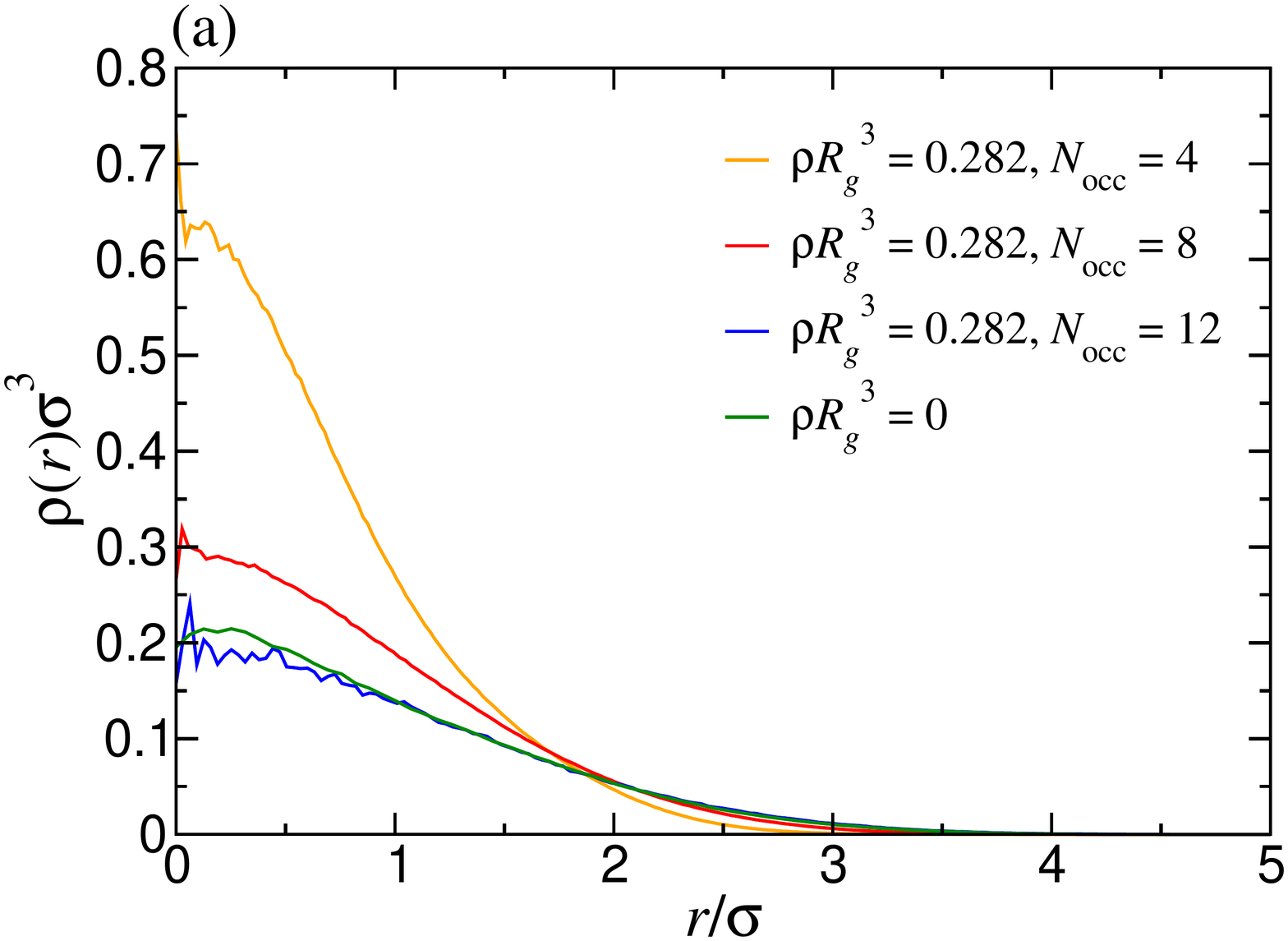}
  \includegraphics[width=8cm]{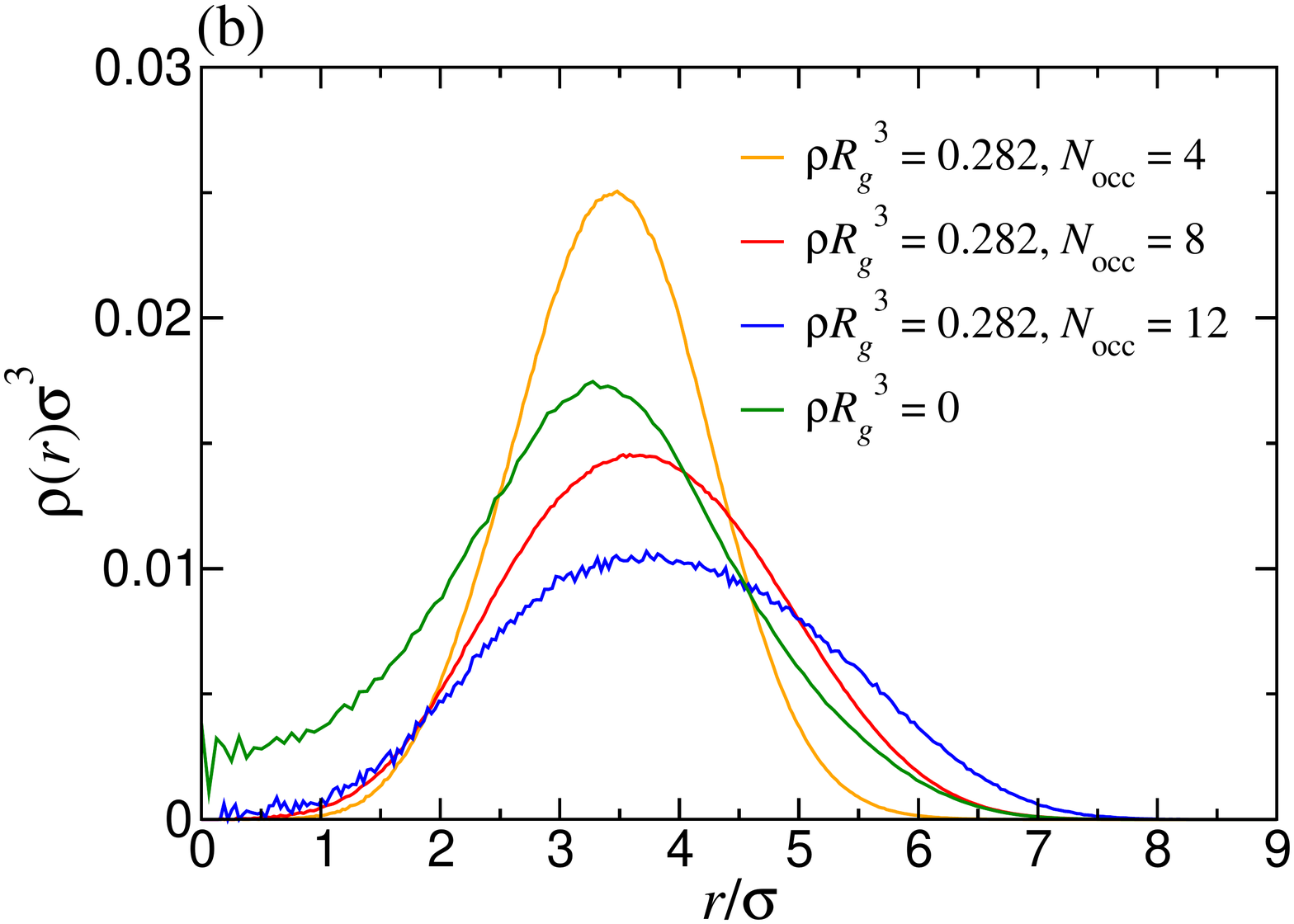}
  \caption{Comparison of the monomer density profiles of dendrimers in the cluster fcc-solid for fixed
  density $\rho$ and varying occupancy $N_{\rm occ}$, as indicated in the legend. As a reference, the
  monomer profiles of isolated amphiphilic dendrimers ($\rho R_g^3 = 0$) are also shown. (a) Core profiles;
  (b) Shell profiles.}
  \label{Fig:clustercrystal:DiC_structure}
\end{figure*}

The fact that $R_g(\rho)$ turns out to be also almost identical to its
zero-density value does not imply, however, that the {\it
  conformation} of the dendrimer is invariant with density. In
Fig.~\ref{Fig:clustercrystal:DiC_structure}, the monomer density
profiles with respect to the dendrimer's center or mass in cluster
crystals are shown. In particular, in
Fig.~\ref{Fig:clustercrystal:DiC_structure}(a) the profiles of the
core monomers ($g = 0$ and $g = 1$) are compared for several
occupations $N_{\text{occ}}$ at fixed density $\rho R_g^3=0.282$,
including the profiles for the equilibrium occupancy, $N_{\rm
  occ}^{\rm eq} \cong 8$; as a reference, the profile of an isolated
dendrimer is also plotted. The same comparison is shown for the shell
monomers ($g = 2$) in Fig.~\ref{Fig:clustercrystal:DiC_structure}(b).
We find that for all finite-density profiles, and in contrast to the
case of the isolated dendrimer, the back-folding of shell monomers to
the center of the dendrimer is suppressed due to the presence of core
monomers of other dendrimers which occupy the available
space. Further, increasing the occupancy naturally leads to an
increasing spatial extension of the core-monomer aggregate as well as
an increased lattice constant. Thus, the shell monomer distribution
broadens for increasing occupancy, and the maximum probability
distance moves slightly towards the exterior of dendrimers. It can be
seen that the profiles of the equilibrium system ($N_{\rm occ}^{\rm
  eq} \cong 8$) and of an isolated dendrimer differ significantly from
one another, while their radii of gyration were found to almost
coincide. This equality originates in two opposing effects, namely a
reduced probability of finding shell monomers within the dendrimer
center at finite densities, paired with an increased probability of
finding core monomers at the center.

%\bibliographystyle{apsrev}
%\bibliography{manuscript}

\begin{thebibliography}{28}
\expandafter\ifx\csname natexlab\endcsname\relax\def\natexlab#1{#1}\fi
\expandafter\ifx\csname bibnamefont\endcsname\relax
  \def\bibnamefont#1{#1}\fi
\expandafter\ifx\csname bibfnamefont\endcsname\relax
  \def\bibfnamefont#1{#1}\fi
\expandafter\ifx\csname citenamefont\endcsname\relax
  \def\citenamefont#1{#1}\fi
\expandafter\ifx\csname url\endcsname\relax
  \def\url#1{\texttt{#1}}\fi
\expandafter\ifx\csname urlprefix\endcsname\relax\def\urlprefix{URL }\fi
\providecommand{\bibinfo}[2]{#2}
\providecommand{\eprint}[2][]{\url{#2}}

\bibitem[{\citenamefont{Klein et~al.}(1994)\citenamefont{Klein, Gould, Ramos,
  Clejan, and Mel'cuk}}]{Klein94}
\bibinfo{author}{\bibfnamefont{W.}~\bibnamefont{Klein}},
  \bibinfo{author}{\bibfnamefont{H.}~\bibnamefont{Gould}},
  \bibinfo{author}{\bibfnamefont{R.~A.} \bibnamefont{Ramos}},
  \bibinfo{author}{\bibfnamefont{I.}~\bibnamefont{Clejan}}, \bibnamefont{and}
  \bibinfo{author}{\bibfnamefont{A.~I.} \bibnamefont{Mel'cuk}},
  \bibinfo{journal}{Physica A} \textbf{\bibinfo{volume}{205}},
  \bibinfo{pages}{738} (\bibinfo{year}{1994}).

\bibitem[{\citenamefont{Likos et~al.}(2001)\citenamefont{Likos, Lang,
  Watzlawek, and L{\"o}wen}}]{Likos01}
\bibinfo{author}{\bibfnamefont{C.~N.} \bibnamefont{Likos}},
  \bibinfo{author}{\bibfnamefont{A.}~\bibnamefont{Lang}},
  \bibinfo{author}{\bibfnamefont{M.}~\bibnamefont{Watzlawek}},
  \bibnamefont{and}
  \bibinfo{author}{\bibfnamefont{H.}~\bibnamefont{L{\"o}wen}},
  \bibinfo{journal}{Phys. Rev. E} \textbf{\bibinfo{volume}{63}},
  \bibinfo{pages}{031206} (\bibinfo{year}{2001}).

\bibitem[{\citenamefont{Mladek et~al.}(2006)\citenamefont{Mladek, Gottwald,
  Kahl, Neumann, and Likos}}]{Mladek06}
\bibinfo{author}{\bibfnamefont{B.~M.} \bibnamefont{Mladek}},
  \bibinfo{author}{\bibfnamefont{D.}~\bibnamefont{Gottwald}},
  \bibinfo{author}{\bibfnamefont{G.}~\bibnamefont{Kahl}},
  \bibinfo{author}{\bibfnamefont{M.}~\bibnamefont{Neumann}}, \bibnamefont{and}
  \bibinfo{author}{\bibfnamefont{C.~N.} \bibnamefont{Likos}},
  \bibinfo{journal}{Phys.\ Rev.\ Lett.} \textbf{\bibinfo{volume}{96}},
  \bibinfo{pages}{045701} (\bibinfo{year}{2006}).

\bibitem[{\citenamefont{Sciortino et~al.}(2004)\citenamefont{Sciortino, Mossa,
  Zaccarelli, and Tartaglia}}]{Sci04}
\bibinfo{author}{\bibfnamefont{F.}~\bibnamefont{Sciortino}},
  \bibinfo{author}{\bibfnamefont{S.}~\bibnamefont{Mossa}},
  \bibinfo{author}{\bibfnamefont{E.}~\bibnamefont{Zaccarelli}},
  \bibnamefont{and}
  \bibinfo{author}{\bibfnamefont{P.}~\bibnamefont{Tartaglia}},
  \bibinfo{journal}{Phys. Rev. Lett.} \textbf{\bibinfo{volume}{93}},
  \bibinfo{pages}{055701} (\bibinfo{year}{2004}).

\bibitem[{\citenamefont{{H. Sedgwick} et~al.}(2005)\citenamefont{{H. Sedgwick},
  {K. Kroy}, {A. Salonen}, {M.B. Robertson}, {S.U. Egelhaaf}, and {W.C.K.
  Poon}}}]{Sedgwick05}
\bibinfo{author}{\bibnamefont{{H. Sedgwick}}},
  \bibinfo{author}{\bibnamefont{{K. Kroy}}}, \bibinfo{author}{\bibnamefont{{A.
  Salonen}}}, \bibinfo{author}{\bibnamefont{{M.B. Robertson}}},
  \bibinfo{author}{\bibnamefont{{S.U. Egelhaaf}}}, \bibnamefont{and}
  \bibinfo{author}{\bibnamefont{{W.C.K. Poon}}}, \bibinfo{journal}{Eur. Phys.
  J. E} \textbf{\bibinfo{volume}{16}}, \bibinfo{pages}{77}
  (\bibinfo{year}{2005}).

\bibitem[{\citenamefont{Mladek et~al.}(2008{\natexlab{a}})\citenamefont{Mladek,
  Charbonneau, Likos, Frenkel, and Kahl}}]{Mladek08a}
\bibinfo{author}{\bibfnamefont{B.~M.} \bibnamefont{Mladek}},
  \bibinfo{author}{\bibfnamefont{P.}~\bibnamefont{Charbonneau}},
  \bibinfo{author}{\bibfnamefont{C.~N.} \bibnamefont{Likos}},
  \bibinfo{author}{\bibfnamefont{D.}~\bibnamefont{Frenkel}}, \bibnamefont{and}
  \bibinfo{author}{\bibfnamefont{G.}~\bibnamefont{Kahl}}, \bibinfo{journal}{J.
  Phys.: Condens. Matter} \textbf{\bibinfo{volume}{20}},
  \bibinfo{pages}{494245} (\bibinfo{year}{2008}{\natexlab{a}}).

\bibitem[{\citenamefont{Mladek et~al.}(2008{\natexlab{b}})\citenamefont{Mladek,
  Kahl, and Likos}}]{Mladek08}
\bibinfo{author}{\bibfnamefont{B.~M.} \bibnamefont{Mladek}},
  \bibinfo{author}{\bibfnamefont{G.}~\bibnamefont{Kahl}}, \bibnamefont{and}
  \bibinfo{author}{\bibfnamefont{C.~N.} \bibnamefont{Likos}},
  \bibinfo{journal}{Phys.\ Rev.\ Lett.} \textbf{\bibinfo{volume}{100}},
  \bibinfo{pages}{028301} (\bibinfo{year}{2008}{\natexlab{b}}).

\bibitem[{\citenamefont{Narros et~al.}(2010)\citenamefont{Narros, Moreno, and
  Likos}}]{Narros10}
\bibinfo{author}{\bibfnamefont{A.}~\bibnamefont{Narros}},
  \bibinfo{author}{\bibfnamefont{A.~J.} \bibnamefont{Moreno}},
  \bibnamefont{and} \bibinfo{author}{\bibfnamefont{C.~N.} \bibnamefont{Likos}},
  \bibinfo{journal}{Soft Matter} \textbf{\bibinfo{volume}{6}},
  \bibinfo{pages}{2435} (\bibinfo{year}{2010}).

\bibitem[{\citenamefont{Lenz et~al.}(2011)\citenamefont{Lenz, Mladek, Likos,
  Kahl, and Blaak}}]{Lenz11}
\bibinfo{author}{\bibfnamefont{D.~A.} \bibnamefont{Lenz}},
  \bibinfo{author}{\bibfnamefont{B.~M.} \bibnamefont{Mladek}},
  \bibinfo{author}{\bibfnamefont{C.~N.} \bibnamefont{Likos}},
  \bibinfo{author}{\bibfnamefont{G.}~\bibnamefont{Kahl}}, \bibnamefont{and}
  \bibinfo{author}{\bibfnamefont{R.}~\bibnamefont{Blaak}},
  \bibinfo{journal}{J.\ Phys.\ Chem.\ B} \textbf{\bibinfo{volume}{115}},
  \bibinfo{pages}{7218} (\bibinfo{year}{2011}).

\bibitem[{\citenamefont{Goerbig et~al.}(2003)\citenamefont{Goerbig, Lederer,
  and Smith}}]{Goerbig03}
\bibinfo{author}{\bibfnamefont{M.~O.} \bibnamefont{Goerbig}},
  \bibinfo{author}{\bibfnamefont{P.}~\bibnamefont{Lederer}}, \bibnamefont{and}
  \bibinfo{author}{\bibfnamefont{C.}~\bibnamefont{Morais Smith}},
  \bibinfo{journal}{Phys. Rev. B} \textbf{\bibinfo{volume}{68}},
  \bibinfo{pages}{241302} (\bibinfo{year}{2003}).

\bibitem[{\citenamefont{Cooper}(2008)}]{Cooper08}
\bibinfo{author}{\bibfnamefont{N.}~\bibnamefont{Cooper}},
  \bibinfo{journal}{Adv.\ Phys.} \textbf{\bibinfo{volume}{57}},
  \bibinfo{pages}{539} (\bibinfo{year}{2008}).

\bibitem[{\citenamefont{Saccani et~al.}(2011)\citenamefont{Saccani, Moroni, and
  Boninsegni}}]{Saccani11}
\bibinfo{author}{\bibfnamefont{S.}~\bibnamefont{Saccani}},
  \bibinfo{author}{\bibfnamefont{S.}~\bibnamefont{Moroni}}, \bibnamefont{and}
  \bibinfo{author}{\bibfnamefont{M.}~\bibnamefont{Boninsegni}},
  \bibinfo{journal}{Phys. Rev. B} \textbf{\bibinfo{volume}{83}},
  \bibinfo{pages}{092506} (\bibinfo{year}{2011}).

\bibitem[{\citenamefont{Saccani et~al.}(2012)\citenamefont{Saccani, Moroni, and
  Boninsegni}}]{saccani:prl:2012}
\bibinfo{author}{\bibfnamefont{S.}~\bibnamefont{Saccani}},
  \bibinfo{author}{\bibfnamefont{S.}~\bibnamefont{Moroni}}, \bibnamefont{and}
  \bibinfo{author}{\bibfnamefont{M.}~\bibnamefont{Boninsegni}},
  \bibinfo{journal}{Phys. Rev. Lett.} \textbf{\bibinfo{volume}{108}},
  \bibinfo{pages}{175301} (\bibinfo{year}{2012}).

\bibitem[{\citenamefont{Moreno and Likos}(2007)}]{Moreno07}
\bibinfo{author}{\bibfnamefont{A.~J.} \bibnamefont{Moreno}} \bibnamefont{and}
  \bibinfo{author}{\bibfnamefont{C.~N.} \bibnamefont{Likos}},
  \bibinfo{journal}{Phys.\ Rev.\ Lett.} \textbf{\bibinfo{volume}{99}},
  \bibinfo{pages}{107801} (\bibinfo{year}{2007}).

\bibitem[{\citenamefont{Coslovich et~al.}(2011)\citenamefont{Coslovich,
  Strauss, and Kahl}}]{Coslovich11}
\bibinfo{author}{\bibfnamefont{D.}~\bibnamefont{Coslovich}},
  \bibinfo{author}{\bibfnamefont{L.}~\bibnamefont{Strauss}}, \bibnamefont{and}
  \bibinfo{author}{\bibfnamefont{G.}~\bibnamefont{Kahl}},
  \bibinfo{journal}{Soft Matter} \textbf{\bibinfo{volume}{7}},
  \bibinfo{pages}{2127} (\bibinfo{year}{2011}).

\bibitem[{\citenamefont{Mladek et~al.}(2007)\citenamefont{Mladek, Charbonneau,
  and Frenkel}}]{Mladek07b}
\bibinfo{author}{\bibfnamefont{B.~M.} \bibnamefont{Mladek}},
  \bibinfo{author}{\bibfnamefont{P.}~\bibnamefont{Charbonneau}},
  \bibnamefont{and} \bibinfo{author}{\bibfnamefont{D.}~\bibnamefont{Frenkel}},
  \bibinfo{journal}{Phys.\ Rev.\ Lett.} \textbf{\bibinfo{volume}{99}},
  \bibinfo{pages}{235702} (\bibinfo{year}{2007}).

\bibitem[{\citenamefont{Nikoubashman et~al.}(2012)\citenamefont{Nikoubashman,
  Kahl, and Likos}}]{Nikoubashman12}
\bibinfo{author}{\bibfnamefont{A.}~\bibnamefont{Nikoubashman}},
  \bibinfo{author}{\bibfnamefont{G.}~\bibnamefont{Kahl}}, \bibnamefont{and}
  \bibinfo{author}{\bibfnamefont{C.~N.} \bibnamefont{Likos}},
  \bibinfo{journal}{Soft Matter} \textbf{\bibinfo{volume}{7}},
  \bibinfo{pages}{4121} (\bibinfo{year}{2012}).

\bibitem[{\citenamefont{Neuhaus and Likos}(2011)}]{Neuhaus11}
\bibinfo{author}{\bibfnamefont{T.}~\bibnamefont{Neuhaus}} \bibnamefont{and}
  \bibinfo{author}{\bibfnamefont{C.~N.} \bibnamefont{Likos}},
  \bibinfo{journal}{J.\ Phys.: Condens.\ Matter} \textbf{\bibinfo{volume}{23}},
  \bibinfo{pages}{234112} (\bibinfo{year}{2011}).

\bibitem[{\citenamefont{Zhang et~al.}(2010)\citenamefont{Zhang, Charbonneau,
  and Mladek}}]{Zhang10}
\bibinfo{author}{\bibfnamefont{K.}~\bibnamefont{Zhang}},
  \bibinfo{author}{\bibfnamefont{P.}~\bibnamefont{Charbonneau}},
  \bibnamefont{and} \bibinfo{author}{\bibfnamefont{B.~M.}
  \bibnamefont{Mladek}}, \bibinfo{journal}{Phys. Rev. Lett.}
  \textbf{\bibinfo{volume}{105}}, \bibinfo{pages}{245701}
  (\bibinfo{year}{2010}).

\bibitem[{\citenamefont{Likos et~al.}(2007)\citenamefont{Likos, Mladek,
  Gottwald, and Kahl}}]{Likos07}
\bibinfo{author}{\bibfnamefont{C.~N.} \bibnamefont{Likos}},
  \bibinfo{author}{\bibfnamefont{B.~M.} \bibnamefont{Mladek}},
  \bibinfo{author}{\bibfnamefont{D.}~\bibnamefont{Gottwald}}, \bibnamefont{and}
  \bibinfo{author}{\bibfnamefont{G.}~\bibnamefont{Kahl}}, \bibinfo{journal}{J.\
  Chem.\ Phys.} \textbf{\bibinfo{volume}{126}}, \bibinfo{pages}{224502}
  (\bibinfo{year}{2007}).

\bibitem[{\citenamefont{Dautenhahn and Hall}(1994)}]{Dautenhahn94}
\bibinfo{author}{\bibfnamefont{J.}~\bibnamefont{Dautenhahn}} \bibnamefont{and}
  \bibinfo{author}{\bibfnamefont{C.~K.} \bibnamefont{Hall}},
  \bibinfo{journal}{Macromolecules} \textbf{\bibinfo{volume}{27}},
  \bibinfo{pages}{5399} (\bibinfo{year}{1994}).

\bibitem[{\citenamefont{Louis et~al.}(2000)\citenamefont{Louis, Bolhuis,
  Hansen, and Meijer}}]{Louis00}
\bibinfo{author}{\bibfnamefont{A.~A.} \bibnamefont{Louis}},
  \bibinfo{author}{\bibfnamefont{P.~G.} \bibnamefont{Bolhuis}},
  \bibinfo{author}{\bibfnamefont{J.~P.} \bibnamefont{Hansen}},
  \bibnamefont{and} \bibinfo{author}{\bibfnamefont{E.~J.}
  \bibnamefont{Meijer}}, \bibinfo{journal}{Phys.\ Rev.\ Lett.}
  \textbf{\bibinfo{volume}{85}}, \bibinfo{pages}{2522} (\bibinfo{year}{2000}).

\bibitem[{\citenamefont{Capone et~al.}(2010)\citenamefont{Capone, Hansen, and
  Coluzza}}]{Capone10}
\bibinfo{author}{\bibfnamefont{B.}~\bibnamefont{Capone}},
  \bibinfo{author}{\bibfnamefont{J.-P.} \bibnamefont{Hansen}},
  \bibnamefont{and} \bibinfo{author}{\bibfnamefont{I.}~\bibnamefont{Coluzza}},
  \bibinfo{journal}{Soft Matter} \textbf{\bibinfo{volume}{6}},
  \bibinfo{pages}{6075} (\bibinfo{year}{2010}).

\bibitem[{Note1()}]{Note1}
Note1, \bibinfo{note}{see Supplemental Material for
  details on the dendrimer model chosen here.}

\bibitem[{Note2()}]{Note2}
Note2, \bibinfo{note}{see Supplemental Material (Fig. S1) for the general shape of the effective interactions of amphiphilic
  dendrimers.}




\bibitem{LenzXX}
\bibinfo{author}{\bibfnamefont{D.~A.} \bibnamefont{Lenz}}, \bibinfo{author}{\bibfnamefont{B.~M.} \bibnamefont{Mladek}}, \bibinfo{author}{\bibfnamefont{C.~N.} \bibnamefont{Likos}},  \bibnamefont{and} \bibinfo{author}{\bibfnamefont{R.}~\bibnamefont{Blaak}}, \bibnamefont{(to be published)}.

\bibitem[{\citenamefont{Alexander and McTague}(1978)}]{Alexander78}
\bibinfo{author}{\bibfnamefont{S.}~\bibnamefont{Alexander}} \bibnamefont{and}
  \bibinfo{author}{\bibfnamefont{J.}~\bibnamefont{McTague}},
  \bibinfo{journal}{Phys. Rev. Lett.} \textbf{\bibinfo{volume}{41}},
  \bibinfo{pages}{702} (\bibinfo{year}{1978}).

\bibitem[{\citenamefont{Van~Santen}(1984)}]{Ostwald84}
\bibinfo{author}{\bibfnamefont{R.~A.} \bibnamefont{Van~Santen}},
  \bibinfo{journal}{J. Phys. Chem.} \textbf{\bibinfo{volume}{88}},
  \bibinfo{pages}{5768} (\bibinfo{year}{1984}).

\bibitem[{Note3()}]{Note3}
Note3, \bibinfo{note}{see Supplemental Material for
  details on this behavior.}

\end{thebibliography}

\begin{thebibliography}{4}
\expandafter\ifx\csname natexlab\endcsname\relax\def\natexlab#1{#1}\fi
\expandafter\ifx\csname bibnamefont\endcsname\relax
  \def\bibnamefont#1{#1}\fi
\expandafter\ifx\csname bibfnamefont\endcsname\relax
  \def\bibfnamefont#1{#1}\fi
\expandafter\ifx\csname citenamefont\endcsname\relax
  \def\citenamefont#1{#1}\fi
\expandafter\ifx\csname url\endcsname\relax
  \def\url#1{\texttt{#1}}\fi
\expandafter\ifx\csname urlprefix\endcsname\relax\def\urlprefix{URL }\fi
\providecommand{\bibinfo}[2]{#2}
\providecommand{\eprint}[2][]{\url{#2}}

\bibitem[{\citenamefont{Mladek et~al.}(2008)\citenamefont{Mladek, Kahl, and
  Likos}}]{Mladek08}
\bibinfo{author}{\bibfnamefont{B.~M.} \bibnamefont{Mladek}},
  \bibinfo{author}{\bibfnamefont{G.}~\bibnamefont{Kahl}}, \bibnamefont{and}
  \bibinfo{author}{\bibfnamefont{C.~N.} \bibnamefont{Likos}},
  \bibinfo{journal}{Phys.\ Rev.\ Lett.} \textbf{\bibinfo{volume}{100}},
  \bibinfo{pages}{028301} (\bibinfo{year}{2008}).

\bibitem[{\citenamefont{Lenz et~al.}(2011)\citenamefont{Lenz, Mladek, Likos,
  Kahl, and Blaak}}]{Lenz11}
\bibinfo{author}{\bibfnamefont{D.~A.} \bibnamefont{Lenz}},
  \bibinfo{author}{\bibfnamefont{B.~M.} \bibnamefont{Mladek}},
  \bibinfo{author}{\bibfnamefont{C.~N.} \bibnamefont{Likos}},
  \bibinfo{author}{\bibfnamefont{G.}~\bibnamefont{Kahl}}, \bibnamefont{and}
  \bibinfo{author}{\bibfnamefont{R.}~\bibnamefont{Blaak}},
  \bibinfo{journal}{J.\ Phys.\ Chem.\ B} \textbf{\bibinfo{volume}{115}},
  \bibinfo{pages}{7218} (\bibinfo{year}{2011}).

\bibitem[{\citenamefont{Welch and Muthukumar}(1998)}]{Welch98}
\bibinfo{author}{\bibfnamefont{P.}~\bibnamefont{Welch}} \bibnamefont{and}
  \bibinfo{author}{\bibfnamefont{M.}~\bibnamefont{Muthukumar}},
  \bibinfo{journal}{Macromolecules} \textbf{\bibinfo{volume}{31}},
  \bibinfo{pages}{5892} (\bibinfo{year}{1998}).

\bibitem[{\citenamefont{Mladek and Frenkel}(2011)}]{Mladek11}
\bibinfo{author}{\bibfnamefont{B.~M.} \bibnamefont{Mladek}} \bibnamefont{and}
  \bibinfo{author}{\bibfnamefont{D.}~\bibnamefont{Frenkel}},
  \bibinfo{journal}{Soft Matter} \textbf{\bibinfo{volume}{7}},
  \bibinfo{pages}{1450} (\bibinfo{year}{2011}).

\end{thebibliography}

\end{document}